\documentclass{eptcs}
\providecommand{\event}{TFPIE 2019}
\usepackage[noxcolor]{pstricks}
\usepackage{mweights}
\usepackage{fontaxes}
\usepackage{pst-node}
\usepackage{graphicx}
\usepackage{subfigure}
\usepackage{stfloats}
\usepackage{multido}
\usepackage{setspace}
\usepackage{textcomp}
\usepackage{amssymb}
\usepackage{alltt}
\RequirePackage{doi}
\usepackage{hyperref}

\newcommand{\while}{\texttt{while} loops \ }
\newcommand{\whilen}{\texttt{while} loops}
\newcommand{\htdp}{\texttt{HtDP} \space}
\newcommand{\htdpn}{\texttt{HtDP}}
\newcommand{\shu}{\texttt{SHU} \space}
\newcommand{\shun}{\texttt{SHU}}

\newcommand{\racketn}{\texttt{Racket}}

\title{How to Design \emph{while} Loops}

\author{Marco T. Moraz\'an
\institute{Seton Hall University}
\email{morazanm@shu.edu}}
\def\titlerunning{How to Design \emph{while} Loops}
\def\authorrunning{M. T. Moraz\'an}

\begin{document}
\maketitle

\begin{abstract}
Beginning students find the syntactic construct known as a \texttt{while} loop difficult to master. The difficulties revolve around guaranteeing loop termination and around learning how to properly sequence mutations to solve a problem. In fact, both of these are intertwined and students need to be taught a model that helps them reason about how to design \whilen. For students that have been introduced to how to design programs using structural recursion, generative recursion, accumulative recursion, and mutation, the task of teaching them how to design \while is made easier. These students are familiar, for example, with state variables, termination arguments, and accumulator invariants. All of these are fundamental in the design of \whilen. This articles presents a novel technique used at Seton Hall University to introduce beginners to the design of \whilen. It presents a design recipe that students can follow step-by-step to establish such things as the driver of the loop, the loop invariant, and the proper sequencing of mutations. The article also presents an example of designing a \texttt{while}-loop based function using the new design recipe.
\end{abstract}

\section{Introduction}
Every instructor that has ever introduced beginners to \whilen \ is witness to how students struggle with this language construct. Toy examples of \whilen \ may not pose a major challenge to many students. Once loops increase in complexity, however, student frustration tends to settle in. It is not uncommon for beginning students to feel frustrated with, what to them are inexplainable, infinite loops. Even more frustrating are loops that terminate, but produce the wrong result. Through years of teaching, instructors have heard many times from a myriad of frustrated students ``\emph{my loop runs, but it is not giving me the right result}." The source of the frustration has its roots in the fact that most textbooks (and, therefore, probably most instructors) only teach students the syntax of \whilen \ and then present examples (e.g., \cite{Ford,Goodrich,Sedgewick,Tymann}). Some textbooks even warn the reader that they must avoid infinite loops. The expectation is that somehow students will learn how to successfully write loops based on syntax, examples, and warnings alone. Although many readers of this article, surely, ``learned" how to write \whilen \ this way, it is time for CS educators to recognize that this approach is utter nonsense. Students, especially beginners, need a systematic way to design \whilen.

At the heart of the teaching-\texttt{while-loops}-problem to beginners is that syntax, examples, and warnings do not provide beginners with a means to understand \emph{why} \whilen \ are infinite or halt or \emph{why} they produce a correct/incorrect result. In other words, learning to program with \whilen \ is hard (as evidenced, for example, by: \cite{Cherenkova,Ginat,Izu,Martin}). These loops are difficult for students, because the syntax hides two important concepts. The first is that \whilen \ are, in essence, syntactic sugar for accumulative recursion--including, accumulative tail-recursion. The second is that \whilen \ are typically used with the sequencing of mutators and/or state-variable mutations. Roughly speaking, this sequencing corresponds to the arguments passed as input to a recursive call made by an accumulative recursive function. If \whilen \ correspond in this manner to accumulative recursive functions, then this means that \whilen \ must be \emph{designed} to exploit accumulators and state variables. It does not suffice to simply teach syntax just like it does not suffice to simply teach syntax for accumulative recursion and for the manipulation of state variables. Students must learn \emph{how} to solve problems using \whilen. An added benefit of designing \whilen \ is that it helps students to organize their thoughts and to communicate the solution to a problem to others--a fundamental pillar of programming \cite{PPL}.

As part of the design process for accumulative recursion, students must understand what an accumulator represents and how to exploit it \cite{HtDP,HtDP2}. That is, they must understand how to use accumulators to solve problems. In essence, this means that they must develop an invariant that must be true every time an accumulative recursive function is called. Interestingly enough, an invariant ought to be developed as part of designing a \texttt{while} loop. Unfortunately, this is rarely done or taught in class leaving students to discover their own ad hoc means to write loops. Students that have learned to design accumulative recursive functions, however, can be shown that the \texttt{while} construct is syntactic sugar for such functions. Thus, students can use this knowledge as a basis to learn how to design while loops. The parameters of an accumulative recursive function are represented as state variables and students must determine the proper sequencing of mutations to write a \texttt{while} loop's body. This second task is nontrivial and is mostly unrelated to solving the problem (i.e., accumulative recursion does not require variable mutation). This means that students must not only solve a problem, but also have the added burden of finding the right sequencing of mutations. To accomplish all this students need a framework that allows them to reason about how to design \whilen. This framework needs to provide them with the tools to identify the role of accumulators as well as to identify the right order in which to mutate state variables. This can be accomplished by teaching students rudimentary denotational semantics to exploit a loop invariant to design their code. This is important, because programmers that identify and write down invariants are more likely to write correct code \cite{Scott}, thus, eliminating a major source of student frustration.

This article presents such a framework in the form of a \emph{design recipe} in the style put forth in the textbook \emph{How to Design Programs} (\htdpn) \cite{HtDP,HtDP2}. In addition, it presents how, at Seton Hall University (\shun), an effective transition is made from designing accumulative recursive functions to the design of \whilen. The article is organized as follows. Section \ref{rw} briefly reviews related work. Section \ref{sb} describes the students' background. Section \ref{acc} outlines the design of accumulative recursive functions. Section \ref{mut} outlines the design of mutation-based functions. Section \ref{tr} presents how the new syntax is initially introduced and used with students. Section \ref{dr} presents a design recipe for \whilen. This design recipe builds on lessons from the design of accumulative recursive functions and from mutation-based computations. Students that lack this background can still benefit from the design recipe, but will need to be introduced to accumulators and invariants for accumulators first. Section \ref{ex} presents an extended example of using the design recipe in the classroom. Finally, Section \ref{fw} presents some conclusions and directions for future work.

\section{Related Work}
\label{rw}
Many textbooks describe \whilen \ as the simplest kind of loop (e.g., \cite{Barclay,Goodrich,Harbison}). A \texttt{while} loop is described as testing if a condition holds and if it does the body of the loop is performed. After each execution of the body, the loop condition is retested to determine if the body ought to be executed again. If the condition fails, if ever, the loop is exited and the program continues execution just beyond the body of the loop. Others add to the description of \whilen \ that they are a syntactic structure to handle computations that are inherently repetitive (e.g., \cite{Kelley,Sedgewick}). Such descriptions are mostly operational descriptions of the mechanical execution of \whilen \ and fail to reveal anything about the semantics of loops or the design methodology that must be employed to write \whilen. In contrast, the work in this article delves directly into how to reason about the design of \whilen. It gives beginners a clear methodology for developing the correct sequencing of mutations that conform the body of the loop.

Some textbooks go a step further and discuss the concept of a loop-invariant and of a post-condition to establish the correctness and termination of a loop (e.g., \cite{Tymann}). A loop invariant is described as a condition that holds after every iteration of the loop and the post-condition is described as the condition that forces the loop to exit\footnote{A more proper term for this condition is the negation of the loop's driver.}. These concepts, however, are only given, at best, a lukewarm treatment and are presented as something that is developed \emph{after} the loop has been written. In contrast, for the work presented in this article these concepts are not an afterthought. Loop invariants and post-conditions are used as integral tool for code development. It is precisely the loop invariant that is useful in establishing the correct sequence of mutations in the body of a loop.

The best-known textbook to teach beginners about program by design is \htdp \cite{HtDP,HtDP2}. The methodology used by \htdp is based on the concept of the design recipe. A design recipe is a series of concrete steps--each with a specific outcome--that helps students design solutions to problems. The current edition of \htdpn \ \cite{HtDP2} presents design recipes for structural, generative, and accumulative recursion. In addition, the first edition of \cite{HtDP} also presents design recipes for state-based programming (i.e., using mutations/assignments). The design recipe for generative recursion, for example, has the development of a termination argument as an explicit step. The design recipe for accumulative recursion has the development of accumulator invariants--conditions about the values the accumulators must hold every time a function is called--as an explicit step. The design recipe for state-based computations includes explicit steps on how to describe and use state variables. \htdpn, however, does not cover the design of \whilen. The work presented in this article applies all of the mentioned \htdpn \ concepts to the design and implementation of \whilen. Accumulator invariants, for example, are explicitly made part of loop invariants and termination arguments are part of any \texttt{while}-loop design.

A great deal of work has been done on program semantics, but surprisingly very little of it makes it into programming textbooks for beginners. Given how so many textbooks and programming languages emphasize state-based computing, it is surprising that even the most rudimentary forms of denotational semantics are not addressed. It is really not too difficult for beginning students to understand, for example, basic Hoare logic \cite{Hoare1998}. The work presented in this article borrows techniques from denotational semantics and uses them as tools to design \whilen. Specifically, a loop invariant is ``dragged" through proposed mutations in the body of a loop to ascertain the correct ordering of the mutations. At the beginning of a loop, these mutations make the invariant temporarily false and the remaining mutations must restore the invariant and make progress towards termination. Once the invariant is restored, the body of the loop has been designed.

\section{Student Background}
\label{sb}
At SHU, the first two introductory Computer Science (\texttt{CS}) courses focus on problem solving using a computer \cite{mtm22,mtm24}. The languages of instruction are the successively richer subsets of \texttt{Racket} known as the student languages which are tightly-coupled with \texttt{HtDP} \cite{HtDP,HtDP2}. No prior experience with programming is assumed. The first course starts by familiarizing students with primitive data (e.g., numbers, strings, booleans, symbols, and images), primitive functions, and library functions to manipulate images (i.e., the image teachpack). During this introduction, students are taught about variables, defining their own functions, and the importance of writing contracts and purpose statements. The next step of the course introduces students to data analysis and programming with compound data of finite size (i.e., structures). At this point, students are introduced to the first design recipe. Students gain experience in developing data definitions, examples for data definitions, function templates, and tests for all the functions they write. Building on this experience, students develop expertise with processing compound data of arbitrary size such as lists, natural numbers, and trees. In this part of the course, students learn to design functions using structural recursion. After structural recursion, students are introduced to functional abstraction and the use of higher-order functions such as \texttt{map} and \texttt{filter}. The first semester ends with a module on distributed programming \cite{mtm26,mtm27}.

In the second course, students are exposed to generative recursion, accumulative recursion, and mutation \cite{mtm24,mtm25}. The course starts with generative recursion. At this point, students have their first exposure to the development of termination arguments. After this, the course exposes students to accumulative recursion and iteration (i.e., tail-recursion). During this module, students learn to develop accumulator invariants and to exploit accumulators. The course ends with two modules on mutation. It is at the end of second module on mutation-based programming that students are exposed to the novel techniques regarding \whilen \ presented in this article.

Each course is for a semester (15 weeks). There are two weekly 75-minute lectures that students are required to attend. The typical classroom has between 20 to 25 students. In addition to the lectures, the instructor is available to students during office hours (3 hours/week) and there are 20-30 hours of tutoring each week which the students may voluntarily attend. The tutoring hours are conducted by undergraduate students handpicked and trained by the lecturer. These tutors focus on making sure students develop answers for each step of the design recipe (from writing contracts to running tests). Students must attempt to follow the steps of the design recipe prior to attending tutoring. Based on a student's work, the tutors and the instructor provide guidance but do not solve problems. Students are still responsible for successfully completing all steps of the design recipe. In addition, tutors attend lectures to assist students when they get stuck with, for example, syntax errors. This type of team-teaching with undergraduate tutors has proven to be extremely well-received by students and to be an effective means to enhance the learning experience.

\section{Design of Accumulative Recursion Functions}
\label{acc}
\texttt{CS} students at \shu during their first-year programming course are presented with two motivations for using accumulators. The first is avoid the loss of knowledge. For example, to find a path from \texttt{A} to \texttt{B} in a directed graph, the nodes visited need to be remembered to avoid an infinite recursion caused by loops in the graph. This knowledge (i.e., the nodes visited) is stored in an accumulator. Every recursive call adds the new node processed to the accumulator. The second is to eliminate delayed operations. Eliminating delayed operations is an optimization that may make programs faster--something that seems universally appealing to \texttt{CS} students. In the presence of delayed operations, the evaluation of any argument that applies a function, \texttt{f}, requires two jumps. A jump to evaluate \texttt{f}'s body and a jump to return to the caller to finish the delayed operation. By using accumulators, delayed operations are eliminated and all function calls can be implemented as a single jump requiring no return.

Regardless of the reason to use accumulators, the design recipe for accumulative recursion states that each accumulator must have an invariant and that initializing and exploiting the accumulator must be clearly stated \cite{HtDP,HtDP2}. The accumulator invariant is an assertion about the value the accumulator stores. Every time the function is called, it is the programmer's responsibility to make sure that the argument provided makes the invariant true. The initial call to the function, must provide the initial value of the accumulator and this value must make its invariant true. When the function terminates, the invariant ought to be useful in establishing the correct value to return.

\begin{figure}[t]
\begin{alltt}
               ; fact: natnum \(\rightarrow\) natnum
               ; Purpose: To compute the factorial of the given natnum
               (define (fact n)
                 (cond [(= n 0) 1]
                       [else (* n (fact (- n 1)))]))

               (check-expect (fact 0) 1)
               (check-expect (fact 3) 6)
\end{alltt}
\caption{Factorial Using Structural Recursion.} \label{fact}
\end{figure}

To make this concrete consider the function to compute \texttt{n!} in Figure \ref{fact}. This function is designed by students using structural recursion on natural numbers. Students are explained that the function \texttt{*} is a delayed operation, because its evaluation must wait for the recursive call, \texttt{(fact (- n 1))}, to be resolved. This makes \texttt{fact} a candidate to use an accumulator to eliminate the delayed operation.

\begin{figure}[t]
\begin{alltt}
               ; fact: natnum \(\rightarrow\) natnum
               ; Purpose: To compute the factorial of the given natnum
               (define (fact n)
                 (local [; fact-accum: natnum natnum \(\rightarrow\) natnum
                         ; Purpose: To compute n!
                         ; Accum Inv: accum = \(\Pi\sb\texttt{i=k+1}\sp\texttt{n} i\)
                         (define (fact-accum k accum)
                           (cond [(= k 0) accum]
                                 [else (fact-accum (sub1 k) (* accum k))]))]
                   (fact-accum n 1)))

               (check-expect (fact 0) 1)
               (check-expect (fact 3) 6)
\end{alltt}
\caption{Factorial Using Accumulative Recursion.} \label{fact-accum}
\end{figure}

\begin{figure*}
\begin{alltt}
; fact: natnum \(\rightarrow\) natnum                       ; fact: natnum \(\rightarrow\) natnum
; Purpose: To compute n!                       ; Purpose: To compute n!
(define (fact n)                               (define (fact n)
 (local [                                       (local [
  ; natnum                                        ; natnum
  ; Purpose: The next accum factor                ; Purpose: The next accum factor
  ; Invariant: k \(\geq\) 0                              ; Invariant: k \(\geq\) 0
  (define k (void))                               (define k (void))
  ; natnum                                        ; natnum
  ; Purpose: The product so far                   ; Purpose: The product so far
  ; Invariant: accum=\(\Pi\sb\texttt{i=k+1}\sp\texttt{n} i\)                     ; Invariant: accum=\(\Pi\sb\texttt{i=k+1}\sp\texttt{n} i\)
  (define accum (void))                           (define accum (void))
  (define (fact-state)                            (define (fact-state)
   (cond                                           (cond
    [(= k 0) accum]                                 [(= k 0) accum]
    [else                                           [else
     (begin                                          (begin
      \emph{(set! k (sub1 k))}                               \emph{(set! accum (* k accum))}
      \emph{(set! accum (* k accum))}                        \emph{(set! k (sub1 k))}
      (fact-state))]))]                               (fact-state))]))]
  (begin                                          (begin
     (set! k n)                                     (set! k n)
     (set! accum 1)                                 (set! accum 1)
     (fact-state))))                                (fact-state))))

(check-expect (fact 0) 1)                      (check-expect (fact 0) 1)
(check-expect (fact 3) 6)                      (check-expect (fact 3) 6)
\end{alltt}
\caption{Two Plausible Implementations of Mutation-Based Factorial.} \label{fchoice}
\end{figure*}

To use an accumulator, the students introduce a locally defined function that, in this case, has a natural number, \texttt{k}, and the accumulator, \texttt{accum}, as parameters. Before writing the body of this function, students must identify the accumulator invariant. With the aid of examples, students observe that the value of the product of the natural numbers processed so far must be accumulated. This means the accumulator is a natural number. Students are led to realize that the natural numbers needed to compute \texttt{n!} are partitioned in two. More formally, the accumulator is a natural number that stores the product of all the natural numbers in [k+1..n]: $\Pi_\texttt{i=k+1}^\texttt{n} i$. As the original function in Figure \ref{fact}, the body of the function must have a conditional expression designed around processing a natural number using structural recursion.

Armed with the knowledge of what the accumulator denotes, students must determine how to exploit the accumulator. Students are taught that a good heuristic to determine how to exploit the accumulator is to examine its value when the recursion halts.  In this case, the recursion halts when \texttt{k} is 0. By plugging in $\texttt{k} = 0$ into the accumulator invariant, the students observe:\\
\begin{alltt}
               accum = \(\Pi\sb\texttt{i=k+1}\sp\texttt{n} i \)

                     = \(\Pi\sb\texttt{i=1}\sp\texttt{n} i \)

                     = n!
\end{alltt}
Now, the students know how useful the invariant is to determine how to exploit the accumulator. When \texttt{k} is 0, they need to return the accumulator. Otherwise, they must recursively process the value of \texttt{k-1} (structural recursion as before) with a new value for the accumulator that makes the accumulator invariant true. To determine this value, students can rewrite the invariant taking into account the new value of \texttt{k}. This step yields:
\begin{alltt}
               accum = \(\Pi\sb\texttt{i=k+2}\sp\texttt{n} i\)
\end{alltt}
This tells them that the product stored in the accumulator for the recursive call needs to be multiplied by \texttt{k+1}. In the current call, this value is $k$. Therefore, for the recursive call the accumulator must be multiplied by \texttt{k}. The unit tests remain unchanged. The result of this design is displayed in Figure \ref{fact-accum}.

\section{Design of State-Based Functions}
\label{mut}
The program in Figure \ref{fact-accum}, without any delayed operations, is what is called \emph{tail-recursive} or \emph{iterative}. If tail calls are optimized then tail-recursive programs can be evaluated with a constant amount of memory and function calls are implemented as unconditional jumps (i.e., \texttt{GOTO}s). In our example, the function needs memory to store two natural numbers (\texttt{k} and \texttt{accum}). Students are asked to consider the trace of, say, \texttt{(fact 4)}:
\begin{alltt}
                                 k accum
          (fact 4) = (fact-accum 4   1)
                   = (fact-accum 3   4)
                   = (fact-accum 2   12)
                   = (fact-accum 1   24)
                   = (fact-accum 0   24)
                   = 24
\end{alltt}
It is not difficult for students to observe that this is similar to mutating \texttt{k} and \texttt{accum}. Each recursive call mutates \texttt{k} by decrementing it by 1 and mutates \texttt{accum} by multiplying it by \texttt{k}.

Students are explained that the translation to a state-based computation, makes the parameters of \texttt{fact-accum} locally defined state variables. This step is akin to what is know as \emph{registerization} in programming languages \cite{eopl}. To translate the body of the function, the structure remains the same (e.g., the \texttt{if} from \texttt{fact-accum} persists), but a sequence of mutations to change the state variables must be made before calling the new parameterless function. The mutations inside the new local function use the expressions in the recursive calls in the accumulative recursive function. The body of the local function mutates the state variables to store the original values passed in the accumulative recursive version. This leaves students with two equally plausible implementations of factorial. They are displayed in Figure \ref{fchoice}. Each state variable requires an invariant. For \texttt{accum}, the invariant is the same as for the accumulative recursive function. The invariant for \texttt{k} is derived from the sample trace.

The translation from an accumulative recursive function to a state-based function is mostly mechanical. The non-mechanical part of the transformation is the order in which the mutations to the state variables are done in the \texttt{else} branch of the conditional in \texttt{fact-state} (in italics in Figure \ref{fchoice}). Students are asked if \texttt{k} ought to be mutated first or if \texttt{accum} ought to be mutated first. A simple glance of the code by a novice reveals no answers. At this point, students are told that they must use the invariants for each of the state variables to design the code that performs the mutations. The two invariants must be true every time \texttt{fact-state} is called (akin to what is required to call an accumulative recursive function).

For both state-based versions, the initial mutations making \texttt{k = n} and \texttt{accum = 1} must make the two invariants true. Students are shown how to reason to prove this. For example, in class the following reasoning is developed:
\begin{alltt}
          n is a natural number \(\Rightarrow\) n \(\geq\) 0 \(\Rightarrow\) k \(\geq\) 0.

          accum = 1

                = \(\Pi\sb\texttt{i=n+1}\sp\texttt{n} i \), given that n+1>n

                = \(\Pi\sb\texttt{i=k+1}\sp\texttt{n} i \)
\end{alltt}
This establishes that the invariant is achieved. Therefore, it is fine to call \texttt{fact-state}. In class, emphasis is placed on not calling the function if the invariant has not been achieved.

Now the task is too establish if any of the two versions of \texttt{fact-state}, re-establish the invariants before the recursive call. Students are shown how to \emph{drag the invariant} through the mutations in the \texttt{begin}-expression in the \texttt{else} line of the conditional. This approach is essentially introducing students to Hoare logic \cite{Hoare} in order to properly sequence mutations. If we perform this process for the function on the left in Figure \ref{fchoice}, we have:
\begin{alltt}
        ;; k \(>\) 0 \(\wedge\) accum = \(\Pi\sb\texttt{i=k+1}\sp\texttt{n} i\)
        (set! k (sub1 k))
        ;; k \(\geq\) 0 \(\wedge\) accum = \(\Pi\sb\texttt{i=k+2}\sp\texttt{n} i\)
        (set! accum (* k accum))
        ;; k \(\geq\) 0 \(\wedge\) accum = k * \(\Pi\sb\texttt{i=k+2}\sp\texttt{n} i\)
\end{alltt}
Students can observe that the invariant is not established before the recursive call to \texttt{fact-state}. Therefore, this sequence of mutations statements is not correct and the function can not be recursively called.

This process for the function on the right in Figure \ref{fchoice}, yields:
\begin{alltt}
        ;; k \(>\) 0 \(\wedge\) accum = \(\Pi\sb\texttt{i=k+1}\sp\texttt{n} i\)
        (set! accum (* k accum))
        ;; k \(>\) 0 \(\wedge\) accum = \(\Pi\sb\texttt{i=k}\sp\texttt{n} i\)
        (set! k (sub1 k))
        ;; k \(\geq\) 0 \(\wedge\) accum = \(\Pi\sb\texttt{i=k+1}\sp\texttt{n} i\)
\end{alltt}
Students can observe that the invariant is reestablished and, therefore, it is safe to make the recursive call to \texttt{fact-state}.

The final step is to establish that \texttt{n!} is returned by \texttt{fact-state}. Given that the invariant is always true when \texttt{fact-state} is called, it suffices to plug-in 0, the value of \texttt{k} when the recursion terminates, into the invariant to establish that \texttt{accum = n!}. Therefore, the function on the right-hand side of Figure \ref{fchoice} is the correct implementation.

As a homework exercise, students are asked to fix the mutations on the left-hand side of Figure \ref{fchoice} without changing the order in which the state variables are mutated. Tutors report mixed results among students with this exercise. Most students immediately realize that the accumulator ought to be multiplied by \texttt{k+1}, not \texttt{k}, but have difficulty developing the assertions to establish demonstrating that the invariant is restored. They can, however, verbalize (i.e., state in a natural language) correct reasoning. At this point in the course, this is nothing less than a success given that this example is their first exposure to this material.

\section{A New Syntax: \texttt{while}}
\label{tr}

\begin{figure*}[t]
\begin{alltt}
; fact: natnum \(\rightarrow\) natnum
; Purpose: To compute the factorial of the given natnum
(define (fact n)
  (local
    [; natnum
     ; Purpose: The next value to multiply into accum
     ; Invariant: k \(\geq\) 0
     (define k (void))
     ; natnum
     ; Purpose: The value of the product so far
     ; Invariant: accum = \(\Pi\sb\texttt{i=k+1}\sp\texttt{n} i\)
     (define accum (void))
     ; fact-while:   \(\rightarrow\) natnum
     ; Purpose: To compute n!
     (define (fact-while)
       (begin
         (set! k n)
         (set! accum 1)
         ;; Invariant: k \(\geq\) 0 \(\wedge\) accum = \(\Pi\sb\texttt{i=k+1}\sp\texttt{n} i\)
         (while (not (= k 0))
           ;; k \(>\) 0 \(\wedge\) accum = \(\Pi\sb\texttt{i=k+1}\sp\texttt{n} i\)
           (set! accum (* k accum))
           ;; k \(>\) 0 \(\wedge\) accum = \(\Pi\sb\texttt{i=k}\sp\texttt{n} i\)
           (set! k (sub1 k)
           ;; k \(\geq\) 0 \(\wedge\) accum = \(\Pi\sb\texttt{i=k+1}\sp\texttt{n} i\))
         ;; k \(\geq\) 0 \(\wedge\) accum = \(\Pi\sb\texttt{i=k+1}\sp\texttt{n} i\) \(\wedge\) k = 0
         ;; \(\Rightarrow\) accum = \(\Pi\sb\texttt{i=1}\sp\texttt{n} i\) = n!
         accum))]
   (fact-while)))
\end{alltt}
\caption{Factorial Using a \texttt{while} Loop.} \label{fact-while}
\end{figure*}

Students are explained that in many languages it is common to package the repeated mutations of state variables using syntax in which the recursive call is not explicit. Such syntactic constructs are called loops and they come in several variations. For example, \texttt{for}, \texttt{while}, and \texttt{repeat-until} loops. Students are explained that they are all similar and that our focus will be on \texttt{while} loops. From the accumulative recursive state-based implementation of a function, it is simple to explain to students the syntactic transformation to create \whilen. First, the state variables must be initialized to achieve the invariant, as in done in the body of the \texttt{local} of the state-based implementation.

Second, students are asked to clearly identify the conditions under which a recursive call is not made. The conjunction of these conditions is what is called the \emph{driver} of the loop and it controls when the body of the loop is executed. A straightforward way to identify the loop's driver is to negate the conditions that halt the recursion. It is fairly easy for students to see that in our example the driver is \texttt{(not (= k 0))}.

Third, the body of the loop is composed of the code in the recursive cases of the conditional in the accumulative recursive state-based implementation. In our example, the code for the loop's body is easily identified as the expression in the \texttt{else} of the conditional. Given that the code of the \texttt{while}-loop implementation is exactly the same as the code for the state-based implementation, it is easy for students to see that the invariant is the same and holds every time the top of the loop is reached. Students are explained that reaching the top of the loop is tantamount to calling an accumulative recursive function. Therefore, the invariant must hold every time the top of the loop is reached.

\begin{figure}
\begin{enumerate}
  \item Problem Analysis
    \begin{enumerate}
        \item Outline how the problem is solved
        \item Pick a mutable data representation
    \end{enumerate}
  \item Write signature, purpose and effect statements, and function header
  \item Write Tests
  \item Develop the Loop Invariant
  \item Define a function with a local expression as its body
      \begin{enumerate}
        \item Locally declare the state variables as \texttt{(void)}
        \item Define the type and purpose for each state variable
        \item Define headers for helper functions
      \end{enumerate}
  \item Write the body of the local using a \texttt{begin} expression
    \begin{enumerate}
      \item Initialize the state variables to achieve the invariant
      \item Define the while loop
        \begin{enumerate}
          \item Define the driver and write the loop header
          \item Use the invariant to correctly sequence mutations
          \item Make progress towards termination
        \end{enumerate}
      \item Use the negation of the driver and the invariant to determine the value to return
    \end{enumerate}
  \item Develop a Termination Argument
  \item Run Tests
\end{enumerate}
\caption{The Design Recipe for \texttt{while}-Loop-Based Functions.} \label{recipe}
\end{figure}

Fourth, students are explained that the driver being false at the top of the loop is tantamount to the recursive function halting. Therefore, the code after the loop corresponds to the code that is evaluated whenever the state-based accumulative recursion function stops. For our example, it is easy for students to see that the only thing that needs to be done is to return \texttt{accum}.

Figure \ref{fact-while} displays the result of this syntactic transformation for our factorial example\footnote{The syntax for \whilen \ in \texttt{Racket} is implemented using a macro.}. The only task left is to convincingly argue that returning \texttt{accum} is correct. Students are explained that when the loop stops the invariant and the negation of the driver are true. These must be used to determine what value ought to be returned. For our example, \texttt{k = 0} and the invariant are true. As for the state-based accumulative recursive function, plugging in \texttt{k = 0} into the righthand side of the invariant for \texttt{accum} reveals that that \texttt{accum} is \texttt{n!}. Therefore, it is the correct value to return.

At this point students understand the design and implementation of \texttt{fact-while}. This is especially enlightening for students that come with prior programming experience, say, in \texttt{Java} or \texttt{C++}. Suddenly, for these students, writing \whilen \ is an exercise in design rather than an ad hoc trial and error exercise. For students that do not come to the class with such prior programming experience, designing \whilen \ is an extension of the design-based methodology that they have been learning. The important point is that for both groups there is nothing mysterious about \whilen. It is a syntactical construct for state-based accumulative recursion. As such, it is natural for them to now expect a design recipe for \texttt{while} loops.

\section{A Design Recipe for \texttt{while} Loops}
\label{dr}

After seeing in class and doing for homework several exercises to transform an accumulative recursive function into a \texttt{while}-based function, students are presented with the 8-step design recipe presented in Figure \ref{recipe}. This design recipe aims to help students to directly develop \texttt{while}-loop based functions (without going through all the transformations exemplified in Sections \ref{acc}--\ref{tr})\footnote{Students, of course, always have the option to do so.}. The steps have been designed to feel familiar to students building on their experience with other design recipes in \htdpn \ and with the transformation of accumulative recursive functions to \texttt{while}-based functions. In addition, every step in the design recipe has a specific result that helps in the development of any \texttt{while}-loop based function. It is noteworthy that these steps do not only help students write \whilen. The result of each step also assists instructors and tutors to gauge the understanding students have about a problem in order to provide feedback and guidance. Furthermore, the results a student presents for each step of the design recipe can be used to assign grades on homework, quizzes, and exams.

For the first step, students analyze a problem to determine how to solve it. If the students can outline how to solve the problem with a finite number of variables that are repeatedly mutated the same way, then they can consider using a \texttt{while} loop. A useful way to determine this is to successfully trace an example using a table containing the values of the state variables at each step. Some of these state variables will be accumulators, each requiring an accumulator invariant, and others may be the state variables around which the algorithm is designed (e.g., a natural number or a list) from which they can derive the halting condition.

For the second step, students first clearly identify the signature for the function defining the types of the input, if any, and the output. Then they write a purpose statement. If the solution is based on a generative recursion design, students must include a description of how the problem is solved. Finally, students write the function header that contains the same number of parameters as the input in the signature.

For the third step, students write tests to illustrate the behavior of their function. These tests must be thorough and include at least one test for each variety of the data being processed. The tests must also illustrate the effects, if any, of the function. For example, if the function sorts a vector in place then the tests must establish that sample vectors are sorted after applying the function to it.

The fourth step is a (somewhat) novel step for students. The development of the loop invariant is usually the hardest step for beginners. This is the step in which the students establish the relationship between the state variables. If they have used a table in step 1, then they can use the table to identify the variable properties that are true for each row of the table (i.e., true before an iteration of the loop). The invariant must include all the state variables mutated in the loop. Typically, it helps to identify, \emph{P}, what needs to be true after the loop is executed (also called the post-condition) and, \texttt{S}, the stopping condition. If \texttt{P} and \texttt{S} are identified, then the following implication must hold:
\begin{alltt}
          Loop Invariant \(\wedge\) S \(\Rightarrow\) P.
\end{alltt}
When the above implication can be established, the students have, at the very least, a good approximation of the loop invariant.

\begin{figure}
\begin{alltt}
               ; signature:     Purpose:     Effect:
               (define (f-while \ldots)
                 (local [; <type>
                         ; Purpose:
                         (define state-var1 (void))
                            \(\ldots\)
                         ; <type>
                         ; Purpose:
                         (define state-varN (void))
                         <helper functions>]
                   (begin
                     (set! state-var1 …)
                        \(\ldots\)
                     (set! state-varN …)
                     ; <Invariant>
                     (while <driver>
                       <while-body>)
                     ; <Invariant> and (not <driver>)
                     <return value code>))
               ; <Termination argument>
               ) ; closes f-while
               (check-expect (f-while \(\ldots\)) \(\ldots\))
                  \(\ldots\)
               (check-expect (f-while \(\ldots\)) \(\ldots\))
\end{alltt}
\caption{A Template to Design Functions Using \texttt{while} Loops.} \label{template}
\end{figure}

For step five, students use a \texttt{local}-expression as the body of the function. The local declarations must capture the state variables. For each state variable, its type and purpose must be identified. In addition, headers for helper functions should also be defined locally. These helper functions are independently designed and implemented. The design of the current function assumes that the local helper functions satisfy their signature and their purpose. One of the goals of this step is to have students use encapsulation to hide the state variables and to have each function be a complete package containing everything that is needed by the body of the \texttt{local}-expression.

For step 6, the body of the \texttt{local}-expression must be a \texttt{begin}-expression. The first steps in the \texttt{begin}-expression initialize the state variables to achieve the invariant. After initialization, \texttt{while} loop is defined. This step is started by writing the loop header with its driver. The body of the loop is designed by dragging the invariant through the proposed mutations as illustrated in Section \ref{mut}. This means that inside the loop's body the invariant becomes temporarily false and must be restored before the body's end. It is emphasized to students that restoring the invariant does not suffice. As designers, they must also make sure that the mutations make progress towards termination. That is, one or more mutations must bring the computation one step closer to making the driver false. The final design step for the \texttt{begin}-expression is to determine the value that must be returned. Students must use the invariant and the negation of the driver to determine this value.

For step 7, students must argue that the loop terminates. That is, they must explain why their loop eventually halts. If step 6 has been correctly designed, then establishing termination is not too difficult. On the other hand, if the body of the loop fails to always make progress towards termination, then students cannot develop this argument and must re-check their previous steps.

For step 8, students run their tests. If any tests fail, the students must check their work. Tests can fail for two reasons. The first is that there is a bug in their design/code. To remedy this bug, students must check the results of each step of the design recipe and make sure they have correctly performed each step. The second is that a test has been incorrectly written. In this case, they must correctly rewrite the test.

The steps of the design recipe suggest a function template for \texttt{while}-loop-based design. Figure \ref{template} displays this template. Students can use this template as they develop answers for the steps of the design recipe. In this manner, they can piece-wise write functions that use \whilen. Whenever a student is asked or decides to use a \texttt{while}-loop they can use this template as the starting point for the structure of their function. The template emphasizes both correctness (i.e., invariant) and unit testing. This is a sound approach when others, with different levels of expertise, understanding, and needs, may read your code. The unit tests communicate the expected behavior of the function. This usually suffices for most users of the code. The invariant and assertions communicate why the function is correct. This is required if a user of the code needs to guarantee the correctness of the software they are developing.

Finally, it is important to emphasize to students that the steps of the design recipe do more than simply help with code development. They also help to communicate to others how a problem is solved. To demonstrate the importance of this students are asked to think if others would be able to understand and to maintain their code. To make the point concrete, students are asked to update someone else's code. This provides the opportunity to talk to beginners about ethics and professionalism in Computer Science.

\section{The Design Recipe in Action}
\label{ex}

\begin{figure*}
\begin{alltt}
; (vectorof number) \(\rightarrow\) (void)
; Purpose: To sort the given vector in non-decreasing order
; Effect: The given vector elements are rearranged in-place.
(define (ins-vector! V)
  (local
   [; sort!: VINTV\(\sb{V}\)[natnum,natnum] \(\rightarrow\) (void)
    ; Purpose: Sort given vector interval in non-decreasing order
    ; Effect: Given interval elements are rearranged in-place
    (define (sort! low high)
      (local
         [; int
          ; Purpose: Next element index to move to sorted part of V
          (define h (void))]
        (begin
          (set! h high)
          ; INV: V[h+1..high] is sorted in non-decreasing order
          ;      \(\wedge\) h >= low-1
          (while (not (empty-VINTV? low h))
            ; h >= low \(\wedge\) V[h+1..high] is sorted in non-decreasing order
            (insert! h  (sub1 high))
            ; h >= low \(\wedge\) V[h..high] is sorted in non-decreasing order
            (set! h (sub1 h))
            ; h >= low-1 \(\wedge\) V[h+1..high] is sorted in non-decreasing order
          ) ; closes while
          ; h >= low-1 \(\wedge\) V[h+1..high] is sorted in non-decreasing order
          ; \(\wedge\) [low..h] is empty
          ; ==> h < low
          ; ==> h = low-1
          ; ==> V[low..high] is sorted in non-decreasing order
         (void))))

    ; insert!: VINTV\(\sb{V}\)[natnum,natnum] \(\rightarrow\) (void)
    ; Purpose: \(\ldots\)
    ; Effect: \(\ldots\)
    ; Assumption: V[lo+1..hi] is sorted in non-decreasing order
    (define (insert! lo hi)
      \(\ldots\))]
   (sort! 0 (sub1 (vector-length V)))))
(check-expect (begin (ins-vector! (vector)) V) (vector))
(check-expect (begin (ins-vector! (vector 20 76 3 44)) V) (vector 3 44 20 76))
(check-expect (begin (ins-vector! (vector 1 2 3)) V) (vector 1 2 3))
(check-expect (begin (ins-vector! (vector 101 8 87 87 8)) V) (vector 8 8 87 87 101))
 \end{alltt}
 \caption{Outline of In-Place Vector Insertion Sorting.} \label{insort}
\end{figure*}

This section presents an extended example of the design recipe for \whilen \ in action. The example chosen is sorting a vector in place using insertion sorting. Students are familiar with insertion sorting as it has been used as an example to sort a list of numbers using structural recursion earlier in the course. Furthermore, students have also developed an insertion sorting function, designed using structural recursion on vector intervals \cite{VINTV}, to sort a vector of numbers in place. Therefore, this brings a familiar example to the realm of teaching students how to design \texttt{while} loops.

For step 1 of the design recipe, problem analysis in class reveals that a local in-place sorting function, \texttt{sort!}, needs to process a vector interval, \texttt{[low..high]}, for an input vector \texttt{V}. Initially, the entire vector must be sorted, thus, \texttt{sort!} must be called with \texttt{[0..V$_\texttt{len}-1$]}, where \texttt{V$_\texttt{len}$} is the length of \texttt{V}. The sorting function halts when the vector interval is empty. The vector interval can be processed from \texttt{high} to \texttt{low} using a structural recursion approach (that will be implemented using a \texttt{while} loop). At each step, the vector interval will have two parts, the sorted and the unsorted portions, and the high element of the unsorted portion is inserted into the sorted portion. The mutable high index, \texttt{h}, is represented as local integer. An integer, not a natural number, is needed given that eventually the mutable index will become -1 when the vector interval is empty. These observations lead to the design recipe's step 2 signature, purpose statement, effect statement, and function header displayed in Figure \ref{insort}.

For step 3, students are explained that the tests must show that the effect is achieved. In our example, this means that the tests must show that a vector ends in a sorted state. The tests aim to be thorough by demonstrating that the effect is achieved given input vectors with different characteristics: empty, non-empty, sorted, in reversed order, randomly populated, and containing repetitions. The tests are displayed in Figure \ref{insort}.

For step 4, students need to see how to formulate an assertion that captures the idea of the vector interval is divided into the sorted and unsorted portions. After some class discussion, the initial invariant developed is:
\begin{alltt}
     V[low..h] is unsorted \(\wedge\) V[h+1..high] is sorted in non-decreasing order
\end{alltt}
Students are asked if this invariant suffices to prove that the vector is sorted when the vector interval becomes empty. In other words, does the following hold:
\begin{alltt}
     INV \(\wedge\) [low..h] is empty \(\Rightarrow\) V[low..high] is sorted in non-decreasing order
\end{alltt}
During class discussion two conclusions are reached. First, the proposed invariant is insufficient, because it is not possible to determine the value of \texttt{h}. Second, including an assertion for the unsorted portion of \texttt{V} is not useful and, therefore, can be dropped from the invariant. Students observe that when the loop halts, it is necessary for \texttt{h} to be \texttt{low-1}. This leads to the following proposed invariant:
\begin{alltt}
     V[h+1..high] is sorted in non-decreasing order \(\wedge\) h >= low-1
\end{alltt}
At this point, students can argue that when the loop halts \texttt{h=low-1} and, therefore, for the given vector interval the vector is sorted. Thus, this invariant suffices.

For step 5, students locally define \texttt{h} as \texttt{(void)} with the type and purpose displayed in Figure \ref{insort}. The more interesting part of this step is to convince students to take a leap of faith and assume that they can write an in-place function to insert \texttt{V[h]} into the sorted part of the vector. It is not uncommon, however, for some students to remember lessons from designing this function using structural recursion on a vector interval. These students remind the class that \texttt{insert!} must process a vector interval and that all but the first element in the given vector interval are sorted in non-decreasing order. All of these class observations are captured in the local declarations of \texttt{ins-vector!} in Figure \ref{insort}. At this point, it is worth asking students why they have not suggested to define \texttt{insert!} locally inside \texttt{sort!}. Most students state that ``for now" it is less confusing not to ``clutter" \texttt{sort!} with more local definitions. Given that students find it less confusing the author feels that this choice is acceptable. Some students, after digesting the design, do make \texttt{insert!} local to \texttt{sort!}.

For step 6, students quickly realize that \texttt{h} must be initialized to \texttt{high} and that the driver is \texttt{(not (empty-VINTV? low h))}\footnote{\texttt{(define (empty-VINTV? low high) (< high low))}}. Developing the body of the while is carefully guided by the instructor with student input. The first step of to have students realize how the first assertion inside the loop is derived from the invariant and the driver. In our case, this means that \texttt{h $\geq$ low}. Usually after this, the first instinct students have to is to insert \texttt{V[h]} into the sorted part of the vector. It is not uncommon, once again, for students to remember lessons from designing this function using structural recursion on a vector interval and state that the vector interval provided as input to \texttt{insert} cannot contain \texttt{high}. After some class discussion, students feel comfortable stating that after the call to \texttt{insert!}, \texttt{h} has not changed and the sorted vector interval is now \texttt{[h..high]}. This is captured in the assertion after the call to \texttt{insert!} in Figure \ref{insort}. At this point, students can observe that progress has not been made towards termination and that the state variable \texttt{h} has not been mutated. The mutation of \texttt{h} is straightforward for students given that this is a design based on structural recursion. Observing that the invariant is restored is also straightforward for students.

Step 7's termination argument is relatively simple for students, since they know that structural recursion always terminates. All students can state that \texttt{h} starts at \texttt{high}. Every time through the loop \texttt{h} is decremented by 1. Therefore, \texttt{h} will eventually be less than \texttt{low} and the loop terminates.

In order to complete step 8, the \texttt{insert!} function must be implemented. It's design follows in the same spirit of \texttt{sort!}. It presents an excellent opportunity to reinforce the design process for \whilen.

\section{Concluding Remarks and Future Work}
\label{fw}
This article presents a methodology to introduce beginners to \whilen. This methodology is based on a novel design recipe that outlines concrete steps with concrete outcomes to guide students in the design of \whilen. The design recipe is inspired in the translation of accumulative recursive functions into a new syntax in which recursive calls are not explicit and in the work done by the denotational semantics community. The steps of the design recipe presented address two of the biggest difficulties beginners face. The first is how to sequence mutations to solve problems using \whilen. The second is how to design loops that are not infinite and argue that a well-designed loop terminates. An important benefit of the approach is that students are introduced to program correctness. This is useful to them as a new tool in their programming toolbox and serves as a gentle introduction for a course on program verification they may take in the future.

Although the examples in this article use \racketn \ syntax, the methodology is not \racketn \ specific nor it is specific to the \racketn \ student languages. That is, the methodology can be used with many programming languages. Furthermore, the presented design recipe does not depend on \racketn \ in any way. In fact, this methodology is used at Seton Hall University with students in their first year that are taught using the \racketn \ student languages and with students in their second year that are taught using an object-oriented programming language like \texttt{Java}.

Future work includes measuring students' attitudes towards the presented design recipe to determine if they feel it is useful and to determine if it makes them better problem solvers. Future work also includes investigating how to teach students to make \whilen \ iterative (i.e., contain no delayed operations). This is likely achievable by giving students a gentle introduction to the core ideas behind continuation-passing style and by developing a design recipe students can follow to eliminate delayed operations. Finally, future work also includes determining how this methodology can be vertically integrated into the CS curriculum. This problem is very challenging, because many instructors are not familiar with formal methods. The hope is that instructors of higher-level courses will be receptive to the idea given that students come with some background in program correctness.

\section{Acknowledgements}
The author thanks Douglas Troeger from The City College of New York, his Ph.D. advisor, for his introduction to formal methods. The work we did with undergraduates at the time has influenced the development of the design recipe for \whilen. The author also thanks Matthias Felleisen et al. for introducing termination arguments and accumulator invariants in \htdpn. That introduction has proven essential in making the work presented in this article more easily accessible to beginners.

\bibliographystyle{eptcs}
\bibliography{Morazan-How-To-Introduce-While-Loops}
\end{document}